\documentclass[referee]{aa}

\usepackage{natbib}
\bibpunct{(}{)}{;}{a}{}{,} 

\usepackage{graphicx}

\usepackage{txfonts}

\newcommand{\vecg}[1]{\mbox{\boldmath$#1$}}
\newcommand{\ve}[1]{\vecg{#1}}
\def\muas{\hbox{$\mu$as}}

\begin{document}

\title{Light Propagation in the Gravitational Field of Moving Bodies
\\ by means of Lorentz Transformation\\
I. Mass monopoles moving with constant velocities}

\author{Sergei A. Klioner}

\institute{Lohrmann Observatory, Dresden Technical University,
Mommsenstr. 13, 01062 Dresden, Germany}

\offprints{Sergei A. Klioner, \email{klioner@rcs.urz.tu-dresden.de}}

\date{Received \today  / Accepted \today}

\abstract{
We show how to derive the equations of light propagation in the
gravitational field of uniformly moving mass monopoles without
formulating and integrating the differential equations of light
propagation in that field. The well-known equations of light
propagation in the gravitational field of a motionless mass monopole
are combined with a suitable Lorentz transformation. The possibility to
generalize this technique for the more complicated case of uniformly
moving body of arbitrary multipole structure is discussed.
}


\titlerunning{Light Propagation in the Gravitational Fields of Moving Bodies}
\authorrunning{S.A.~Klioner}

\maketitle

\section{Introduction}

Within the next decade the accuracy of space-based astrometric
positional observations is expected to attain an accuracy of 1
microarcsecond (\muas). This technical progress is expected to be
achieved due to a number of space astrometry projects (e.g., GAIA
\citep{GAIA:2000,Perryman:et:al:2001,Bienayme:Turon:2002} and SIM
\citep{SIM:1998} approved by ESA and NASA). Modeling of the observed
data with such an accuracy is a highly non-trivial task that in any
case requires the use of general relativity. The whole reduction scheme
should be consequently formulated within the framework of general
relativity. Recently, a practical relativistic model for positional
observations of microarcsecond accuracy performed from space has been
formulated by \citet{Klioner:2003}. In particular, many subtle
relativistic effects in light propagation should be accounted for to
attain the goal accuracy of 1 \muas. One of the most intricate point of
the whole relativistic model is to compute the effects of the
translational motion of gravitating bodies on the light propagation.

For the first time this problem  was treated probably by
\citet{Hellings:1986} who recommended to use the standard
post-Newtonian formulas for the light propagation in the gravitational
field of a motionless body and substitute in those formulas the
position of each gravitating body at the moment of closest approach of
that body and the photon. Next step has been done by
\citet{Klioner:1989} where the problem has been solved completely for
the bodies moving with a constant velocity in the first post-Newtonian
approximation. The effects of accelerations of the bodies have been
further treated by \citet{Klioner:Kopeikin:1992} where it was shown
that if the coordinates and velocities of the bodies are computed at
the moments of closest approach of the corresponding body and the
photon, the residual terms of the solution are in some sense minimized.
The complete solution of the problem for arbitrarily moving bodies in
the first post-Minkowskian approximation was found by
\citet{Kopeikin:Schaefer:1999} who succeeded to integrate analytically
the post-Minkowskian equations of light propagation in the field of
arbitrarily moving mass monopoles (see Appendix
\ref{appendix:Kopeikin:Schaefer} for an explicit form of the
Kopeikin-Sch\"afer solution). These authors used the representation of
the metric tensor through the Lienard-Wiechert potentials with retarded
argument. The Kopeikin-Sch\"afer scheme has been generalized by
\citet{Kopeikin:Mashhoon:2002} onto the case of arbitrarily moving
bodies possessing mass monopoles and spin dipoles.

Extensive numerical simulations of light propagation in the
time-dependent gravitational field of the solar system have been
recently done by \citet{Klioner:Peip:2003}. That publication contains
also detailed information of all the approaches mentioned above. The
aim of the numerical simulations was to check the practical accuracy of
various approximate analytical solutions for the light propagation in
the field of moving bodies and verify the practical recommendations
formulated by \citet{Klioner:2003}.

Up to now all theoretical results concerning light propagation in the
field of moving bodies concerned moving mass monopoles (i.e. ``point
masses''). It is well known, however, that for GAIA and other space
missions not only the mass monopoles, but also the gravitational fields
produced by higher multipoles (especially, by mass quadrupoles) are
important \citep{Klioner:2003}. This paper is the first one in a series
of papers where the light propagation in the gravitational field of
moving bodies with full multipole structure will be investigated. The
aim of this paper is to show that the Lorentz transformation can be
used to derive the laws of light propagation in the gravitational field
of a system of uniformly moving bodies in the first post-Minkowskian
(or post-Newtonian) approximation of general relativity.

Let us clarify here that the post-Minkowskian approximation scheme
deals with expansions in powers of the gravitational constant $G$. The
first post-Minkowskian approximation implies that all terms of order
${\cal O}(G^2)$ are neglected. The post-Newtonian approximation scheme
operates with expansions in powers of $c^{-1}$. In the first
post-Newtonian approximation terms of order ${\cal O}(c^{-4})$ are
neglected in the equations of light propagation. One can prove that in
the case of light propagation the formulas of the first post-Newtonian
approximation are linear in $G$ and, therefore, contained in those of
the first post-Minkowskian approximation.

It is well known that the differential equations of light propagation
in both the post-Newtonian and post-Minkowskian approximation are
linear with respect to the non-Galilean components of the metric tensor
$h_{\alpha\beta}$:

\begin{equation}\label{g-eta-h}
g_{\alpha\beta}=\eta_{\alpha\beta}+h_{\alpha\beta},
\end{equation}

\noindent
with

\begin{equation}\label{eta-Minkowski}
\eta_{\alpha\beta}={\rm diag}(-1,+1,+1,+1),
\end{equation}

\noindent
i.e. linear with respect to the gravitational constant $G$. It is also
clear that in this approximation the gravitational potentials
$h_{\alpha\beta}$ can be written as a sum of the contributions coming
from each body of a material system. Therefore, the gravitational
influences of each body on the light propagation can be considered
independently of each other.

Let us first assume that each body moves with a constant velocity
relative to a reference system $(t,x^i)$. Then, in order to derive the
influence of a rectilinearly and uniformly moving body $A$ on the light
propagation one can simply introduce a reference system
$\left(T,X^a\right)_A$ where the body is at rest. Reference systems
$\left(T,X^a\right)_A$ and $(t,x^i)$ can be related to each other by
usual Lorentz transformation. In reference system
$\left(T,X^a\right)_A$ the laws of light propagation are very well
known:  these are the laws of light propagation in the gravitational
field of a motionless body (i.e. mass monopole (Schwarzschild) field or
gravitational field with a given multipole structure).  Then applying
the Lorentz transformation to those known laws of light propagation one
restores the more complicated equations of light propagation in the
gravitational field of uniformly moving body $A$. This very transparent
scheme allows one to re-derive and better understand the formulas for
the light propagation in the field of moving bodies found recently by
\citet{Kopeikin:Schaefer:1999} and \citet{Kopeikin:Mashhoon:2002}, and
also generalize these solutions for the case of uniformly moving bodies
with full multipole structure.

\section{General considerations: uniform motion}

\subsection{Lorentz transformation}

Since each body $A$ in the system can be considered in the same way,
the subscript $A$ and the corresponding sums will be omitted below.
Lorentz transformations between the reference system $(t,x^i)$ and the
reference system $(T,X^a)$ where the body is at rest can be written in
the form (throughout the paper repeated indices imply Einsteinian
summation rule irrespective of their positions)
\begin{eqnarray}\label{Lorentz}
c\,t&=&\Lambda^0_0\,c\,T+\Lambda^0_a\,X^a,
\nonumber\\
x^i&=&\Lambda^i_0\,c\,T+\Lambda^i_a\,X^a,
\end{eqnarray}

\noindent
where
\begin{eqnarray}\label{Lorentz-Lambdas}
\Lambda^0_0&=&\gamma,
\nonumber\\
\Lambda^0_a&=&k^a\,\gamma,
\nonumber\\
\Lambda^i_0&=&k^i\,\gamma,
\nonumber\\
\Lambda^i_a&=&\delta^{ia}+{\gamma^2\over 1+\gamma}\,k^i\,k^a,
\nonumber\\
\gamma&=&\left(1-\ve{k}\,\cdot\,\ve{k}\right)^{-1/2},
\nonumber\\
\ve{k}&=&\ve{v}/c.
\end{eqnarray}

\noindent
Here and below a dot between two coordinate 3-vectors denotes a scalar
product with respect to the Euclidean metric, for example,
$\ve{k}\,\cdot\,\ve{k}=k^a\,k^a$. Velocity $\ve{v}$ is the parameter of
the transformation that will be discussed below. The inverse
transformations read
\begin{eqnarray}\label{inverse-Lorentz}
c\,T&=&\overline\Lambda^0_0\,c\,t+\overline\Lambda^0_i\,x^i,
\nonumber\\
X^a&=&\overline\Lambda^a_0\,c\,t+\overline\Lambda^a_i\,x^i,
\end{eqnarray}

\noindent
where
\begin{eqnarray}\label{inverse-Lorentz-Lambdas}
\overline\Lambda^0_0&=&\gamma,
\nonumber\\
\overline\Lambda^0_i&=&-k^i\,\gamma,
\nonumber\\
\overline\Lambda^a_0&=&-k^a\,\gamma,
\nonumber\\
\overline\Lambda^a_i&=&\delta^{ia}+{\gamma^2\over 1+\gamma}\,k^i\,k^a.
\end{eqnarray}

\subsection{Position of the body}

Let $X^a_A={\rm const}$ be the coordinates of the body in the reference
system $(T,X^a)$. Transforming events $(T,X^a_A)$ into the reference
system $(t,x^i)$ yields

\begin{equation}\label{x_A(t)}
\ve{x}_A(t)=\ve{x}_A(t_A)+\ve{v}_A\,(t-t_A),
\end{equation}

\noindent
where

\begin{equation}\label{v_A(t)}
v_A^i=c\,\Lambda^i_0/\Lambda^0_0=v^i,
\end{equation}

\noindent
and $t_A$ can be chosen arbitrarily. These formulas mean that the body
that is at rest in the reference system $(T,X^a)$ moves with a constant
velocity $\ve{v}_A=\ve{v}$ in the reference system $(t,x^i)$. This
velocity serves as the parameter of the transformations
(\ref{Lorentz})--(\ref{Lorentz-Lambdas}) and
(\ref{inverse-Lorentz})--(\ref{inverse-Lorentz-Lambdas})

\subsection{Unperturbed trajectory of the photon}

The unperturbed trajectory of the photon $X_p^a(T)$ in the reference
system $(T,X^a)$ is a straight line

\begin{equation}\label{X_p(T)-unperturbed}
\ve{X}_p(T)=\ve{X}_{p0}+c\,\ve{\Sigma}\,(T-T_0),
\end{equation}

\noindent
where $\ve{X}_{p0}={\rm const}$ is the position of the photon for
$T=T_0$ and $\ve{\Sigma}$ is the unit direction of the light
propagation ($\ve{\Sigma}\,\cdot\,\ve{\Sigma}=1$). Transforming the
events $(T,X_p^a(T))$ into the reference system $(t,x^i)$ one gets

\begin{equation}\label{x_p(t)-unperturbed}
\ve{x}_p(t)=\ve{x}_{p0}+c\,\ve{\sigma}\,(t-t_0),
\end{equation}

\noindent
where
\begin{eqnarray}\label{t0-xp0}
c\,t_0&=&\Lambda^0_0\,c\,T_0+\Lambda^0_a\,X^a_{p0},
\nonumber\\
x^i_{p0}&=&\Lambda^i_0\,c\,T_0+\Lambda^i_a\,X^a_{p0},
\end{eqnarray}

\noindent
and

\begin{equation}\label{sigma-Sigma}
\sigma^i={\Lambda^i_0+\Lambda^i_a\,\Sigma^a\over
\Lambda^0_0+\Lambda^0_a\,\Sigma^a}.
\end{equation}

\noindent
Therefore, one has $\ve{\sigma}\,\cdot\,\ve{\sigma}=1$. The inverse
transformation of $\ve{\sigma}$ into $\ve{\Sigma}$ has exactly the same
form as (\ref{sigma-Sigma}) but with the corresponding components on
the inverse Lorentz transformation:

\begin{equation}\label{Sigma-sigma}
\Sigma^a={\overline\Lambda^a_0+\overline\Lambda^a_i\,\sigma^i\over
\overline\Lambda^0_0+\overline\Lambda^0_i\,\sigma^i}.
\end{equation}

\subsection{First-order perturbations of the light ray}

Let us write the perturbed trajectory of the photon $X_p^a(T)$ in the
reference system $(T,X^a)$ as

\begin{equation}\label{X_p(T)-perturbed}
\ve{X}_p(T)=\ve{X}_{p0}+c\,\ve{\Sigma}\,(T-T_0)+\Delta \ve{X}_p(T),
\end{equation}

\noindent
with
\begin{eqnarray}\label{Delta-X-a-cond}
\Delta \ve{X}_p(T_0)&=&0,
\nonumber\\
\lim_{T\to-\infty}{d\over dT}\Delta \ve{X}_p(T)&=&0.
\end{eqnarray}

Transforming the events $(T,X_p^a(T))$ as defined by
(\ref{X_p(T)-perturbed}) one gets

\begin{equation}\label{x_p(t)-perturbed}
\ve{x}_p(t)=\ve{x}_{p0}+c\,\ve{\sigma}\,(t-t_0)+\Delta \ve{x}_p(t),
\end{equation}

\noindent
with
\begin{eqnarray}\label{Delta-x-i-cond}
\Delta \ve{x}_p(t_0)&=&0,
\nonumber\\
\lim_{t\to-\infty}{d\over dt}\Delta \ve{x}_p(t)&=&0,
\end{eqnarray}

\noindent
and $\ve{\sigma}$ and $\ve{\Sigma}$ are related to each other by the
same equations (\ref{sigma-Sigma})--(\ref{Sigma-sigma}) as in the
unperturbed case. The light path perturbation in the reference system
$(t,x^i)$ reads

\begin{equation}\label{Delta-x-Delta-X}
\Delta x^i_p(t)=\left(\Lambda^i_a-\sigma^i\,\Lambda^0_a\right)\,
\Delta X^a_p(T).
\end{equation}

It is useful also to derive the corresponding transformation formulas
for the coordinate velocity of the photon. In the reference system
$(T,X^a)$ one gets

\begin{equation}\label{V_p(T)-perturbed}
\ve{V}_p(T)\equiv{d\over dT}\,\ve{X}_p(T)=c\,\ve{\Sigma}+\Delta \ve{V}_p(T),
\end{equation}

\noindent
where

\begin{equation}\label{Delta-V_p(T)-perturbed}
\Delta \ve{V}_p(T)={d\over dT}\,\Delta \ve{X}_p(T).
\end{equation}

\noindent
From (\ref{x_p(t)-perturbed}) one has

\begin{equation}\label{v_p(t)}
\ve{v}_p(t)\equiv{d\over dt}\,\ve{x}_p(t)=c\,\ve{\sigma}+\Delta \ve{v}_p(t),
\end{equation}

\noindent
with

\begin{equation}\label{Delta-v_p(t)}
\Delta \ve{v}_p(t)={d\over dt}\,\Delta \ve{x}_p(t).
\end{equation}

\noindent
The Lorentz transformation gives

\begin{equation}\label{v_p(t)-V_p(T)}
v^i_p=c\,{\Lambda^i_0+\Lambda^i_a\,V^a_p/c\over
\Lambda^0_0+\Lambda^0_a\,V^a_p/c}.
\end{equation}

\noindent
Taking $t\to-\infty$ in (\ref{v_p(t)-V_p(T)}) one restores the relation
between $\ve{\sigma}$ and $\ve{\Sigma}$ given by (\ref{sigma-Sigma}).
Therefore, the relation between $\Delta \ve{v}_p(t)$ and $\Delta
\ve{V}_p(T)$ reads
\begin{eqnarray}\label{Delta-v-Delta-V}
\Delta v^i_p(t)&=&
{1\over \Lambda^0_0+\Lambda^0_b\,V^b_p(T)/c}\,
\left(\Lambda^i_a-\sigma^i\,\Lambda^0_a\right)\,\Delta V^a_p(T)
\nonumber\\
&=&
\left(\overline\Lambda^0_0+\overline\Lambda^0_j\,v^j_p(t)/c\right)\,
\left(\Lambda^i_a-\sigma^i\,\Lambda^0_a\right)\,\Delta V^a_p(T),
\end{eqnarray}

\noindent
which can be simplified to
\begin{eqnarray}\label{Delta-v-Delta-V-pM}
\Delta v^i_p(t)
&=&
\left(\overline\Lambda^0_0+\overline\Lambda^0_j\,\sigma^j\right)\,
\left(\Lambda^i_a-\sigma^i\,\Lambda^0_a\right)\,\Delta V^a_p(T)
+{\cal O}(G^2)
\end{eqnarray}

\noindent
in the first post-Minkowskian approximation. The same equation can be
derived directly from (\ref{Delta-x-Delta-X}).

Now substituting some specific expressions for $\Delta \ve{X}_p(T)$
and/or $\Delta \ve{V}_p(T)$ (e.g. those for the Schwarzschild field or
for the gravitational field with some multipole structure) one can
derive the corresponding perturbations $\Delta \ve{x}_p(t)$ and $\Delta
\ve{v}_p(t)$ without integration of any additional differential
equations of light propagation in the reference system where the body
moves.

\section{Mass monopoles in uniform motion}

The solution for the light propagation in the gravitational field of a
motionless mass monopole in the first post-Minkowskian approximation is
very well known (the post-Newtonian and post-Minkowskian solutions
coincide in this case) and was published e.g. by \citet{Will:1993}. In
our notations one has
\begin{eqnarray}\label{Delta-X-Mono}
\Delta\ve{X}_p(T)&=&
-{2GM\over c^2}\,\biggl(\ve{F}(T)-\ve{F}(T_0)\biggr)+{\cal O}(G^2),
\\ \label{F-Mono}
\ve{F}(T)&=&{\ve{D}\over |\ve{R}|-\ve{\Sigma}\,\cdot\,\ve{R}}-\ve{\Sigma}\,
\log\bigl(|\ve{R}|-\ve{\Sigma}\,\cdot\,\ve{R}\bigr),
\\
\label{Delta-V-Mono}
{1\over c}\,\Delta\ve{V}_p(T)&=&
-{2GM\over c^2}\,
\biggl(
{\ve{D}\over |\ve{R}|\bigl(|\ve{R}|-\ve{\Sigma}\,\cdot\,\ve{R}\bigr)}
+{\ve{\Sigma}\over |\ve{R}|}
\biggr)
+{\cal O}(G^2),
\\
\label{ve-D}
\ve{D}&=&\ve{\Sigma}\times\left(\ve{R}\times\ve{\Sigma}\right),
\\
\label{R}
\ve{R}&=&\ve{X}_{p0}+c\,\ve{\Sigma}\,(T-T_0)-\ve{X}_A,
\end{eqnarray}

\noindent
where $\ve{X}_A$ is the constant position of the body. Note that
$\ve{D}$ is a time-independent vector.

Applying the Lorentz transformations
(\ref{Lorentz})--(\ref{inverse-Lorentz-Lambdas}) to each element of
(\ref{Delta-X-Mono})--(\ref{R}) (the most important parts of this
transformation are given in Appendix
\ref{Appendix-Lorentz-transformed}) and using (\ref{Delta-x-Delta-X})
and (\ref{Delta-v-Delta-V-pM}) one gets
\begin{eqnarray}\label{Delta-x-Mono}
\Delta\ve{x}_p(t)&=&
-{2GM\over c^2}\,\biggl(\ve{f}(t)-\ve{f}(t_0)\biggr)+{\cal O}(G^2),
\\ \label{f-Mono}
\ve{f}(t)&=&
{
\ve{r}-\displaystyle{1\over\theta}\,(\ve{\sigma}\cdot\ve{r})\,(\ve{\sigma}-\ve{k})
\over
\left(
\rho
-
\displaystyle{1\over\gamma\,\theta}\,
\ve{\sigma}\cdot\ve{r}+\gamma\,\ve{k}\cdot\ve{r}
\right)
}
\nonumber\\
&&-\,\gamma\,\left(\ve{\sigma}-\ve{k}\right)\,
\log\left(
\rho
-
\displaystyle{1\over\gamma\,\theta}\,
\ve{\sigma}\cdot\ve{r}+\gamma\,\ve{k}\cdot\ve{r}
\right)
,
\\
\label{Delta-v-Mono}
{1\over c}\,\Delta\ve{v}_p(t)&=&
-{2GM\over c^2}\,\gamma\,\theta\,
\left[
{
\ve{r}-\displaystyle{1\over\theta}\,(\ve{\sigma}\cdot\ve{r})\,(\ve{\sigma}-\ve{k})
\over
\rho\,
\left(
\rho
-
\displaystyle{1\over\gamma\,\theta}\,
\ve{\sigma}\cdot\ve{r}+\gamma\,\ve{k}\cdot\ve{r}
\right)
}
\right.
\nonumber\\
&&
\phantom{-{2GM\over c^2}\,\gamma\,\theta\,\Biggl[}
\left.+\,
{\gamma\,\left(\ve{\sigma}-\ve{k}\right)\over
\rho
}
\right]+{\cal O}(G^2),
\\
\label{ve-rho}
\rho&=&\left(|\ve{r}|^2+\gamma^2\,(\ve{k}\cdot\ve{r})^2\right)^{1/2},
\end{eqnarray}

\noindent
where $\ve{r}$ and $\theta$ are defined by (\ref{r}) and (\ref{theta}),
respectively. Neglecting in (\ref{Delta-x-Mono})--(\ref{ve-rho}) all
the terms of order ${\cal O}(k^2)$ (that is retaining all the terms at
most linear relative to the velocity of the body) we immediately
restore the corresponding formulas derived by
\citet{Klioner:1989,Klioner:1991} and \citet{Klioner:Kopeikin:1992} for
the case of uniformly moving bodies in the first post-Newtonian
approximation.

The equations (\ref{Delta-x-Mono})--(\ref{Delta-v-Mono}) can be
drastically simplified by using the positions $x^i_A(t_*)$ of the
gravitating body at the retarded moments of time $t_*$ and $t_{0*}$
corresponding to the moments $t$ and $t_0$ as described in Appendix
\ref{appendix-retarded}:
\begin{eqnarray}\label{Delta-x-Mono-*}
\Delta\ve{x}_p(t)&=&
-{2GM\over c^2}\,\biggl(\ve{f}_*(t)-\ve{f}_*(t_0)\biggr)+{\cal O}(G^2),
\\ \label{f-Mono-*}
\ve{f}_*(t)&=&
\gamma\left(\theta\,{\ve{d_*}\over p_*}
-(\ve{\sigma}-\ve{k})\,\log\,p_*\right),
\\
\label{Delta-v-Mono-*}
{1\over c}\,\Delta\ve{v}_p(t)&=&
-{2GM\over c^2}\,{\gamma\,\theta\over |\ve{r}_*|-\ve{k}\cdot\ve{r}_*}\,
\nonumber\\
&&\times\,
\biggl(\theta\,{\ve{d}_*\over p_*}
+(2-\theta)\,\ve{\sigma}-2\,\ve{k}\biggr)
+{\cal O}(G^2),
\\
\label{p-*}
p_*&=&|\ve{r}_*|-\ve{\sigma}\cdot\ve{r}_*,
\end{eqnarray}

\noindent
where $\ve{d}_*$ is defined by (\ref{d_*}). Note that
\begin{eqnarray}\label{f-f_*}
\ve{f}(t)&=&\ve{f}_*(t)-\gamma\,\left[
\ve{\sigma}\times(\ve{k}\times\ve{\sigma})-
(\ve{\sigma}-\ve{k})\,\log(\gamma\,\theta)\right],
\end{eqnarray}

\noindent
which implies $\ve{f}(t)-\ve{f}(t_0)=\ve{f}_*(t)-\ve{f}_*(t_0)$ since
the second term on the right-hand side of (\ref{f-f_*}) is
time-independent.

The solution (\ref{Delta-x-Mono-*})--(\ref{Delta-v-Mono-*}) completely
coincides with the solution for moving mass monopoles found by
\citet{Kopeikin:Schaefer:1999} and given in Appendix
\ref{appendix:Kopeikin:Schaefer} in explicit form, provided that in the
Kopeikin-Sch\"afer solution one assumes the body to move with a
constant velocity. Note that if one computes also $\ve{k}=\ve{v}/c$ and
the corresponding quantities $\gamma$ and $\theta$ at the retarded
moment $t_*$ the solution (\ref{Delta-v-Mono-*}) for $\Delta\ve{v}_p$
becomes {\it equivalent} to the Kopeikin-Sch\"afer solution and is,
therefore, correct for an arbitrarily moving mass monopole.

\section{Concluding remarks}

Using the well-known light propagation laws in the Schwarzschild field
in the first post-Minkowskian approximation and the Lorentz
transformation we have derived the post-Minkowskian light propagation
laws in the gravitational field of a uniformly moving mass monopole. In
a similar way one can combine the Lorentz transformation and the light
propagation laws in the gravitational field of a body with arbitrary
multipole structure discussed by \citet{Kopeikin:1997} to derive the
light propagation laws in the gravitational field of a uniformly moving
body with full multipole structure. This will be done in a further
publication.

\appendix

\section{Lorentz transformation of some quantities}
\label{Appendix-Lorentz-transformed}

Applying Lorentz transformations
(\ref{Lorentz})--(\ref{inverse-Lorentz-Lambdas}) to the quantities
appearing in (\ref{Delta-X-Mono})--(\ref{R}) one has
\begin{eqnarray}
\label{tr-Sigma}
\ve{\Sigma}&=&{1\over\gamma\,\theta}\,
\left(\ve{\sigma}-{\gamma\over 1+\gamma}\,
\left(1+\gamma\,\theta\right)\,\ve{k}\right),
\\
\label{theta}
\theta&=&1-\ve{\sigma}\cdot\ve{k},
\\
\label{tr-P-Sigma}
\overline{\ve{\Sigma}}&=&
\gamma\,\left(\ve{\sigma}-\ve{k}\right),
\qquad
\overline{\Sigma}^{\,i}=\left(\Lambda^i_a-\sigma^i\,\Lambda^0_a\right)\,\Sigma^a,
\\
\label{tr-R}
\ve{R}&=&\ve{r}+{\gamma^2\over 1+\gamma}\,(\ve{k}\cdot\ve{r})\,\ve{k},
\\
\label{r}
\ve{r}&=&\ve{x}_{p0}+c\,\ve{\sigma}\,(t-t_0)-\ve{x}_A(t),
\\
\label{tr-abs-R}
|\ve{R}|&=&\left(|\ve{r}|^2+\gamma^2\,(\ve{k}\cdot\ve{r})^2\right)^{1/2},
\\
\label{tr-P-R}
\overline{\ve{R}}&=&
\ve{r}-\gamma^2\,(\ve{k}\cdot\ve{r})\,(\ve{\sigma}-\ve{k}),
\qquad
\overline{R}^i=\left(\Lambda^i_a-\sigma^i\,\Lambda^0_a\right)\,R^a,
\\
\label{tr-Sigma-R}
\ve{\Sigma}\cdot\ve{R}&=&{1\over\gamma\,\theta}\,
\ve{\sigma}\cdot\ve{r}-\gamma\,\ve{k}\cdot\ve{r},
\\
\label{tr-P-D}
\overline{\ve{D}}&=&
\ve{r}
-{1\over\theta}\,(\ve{\sigma}\cdot\ve{r})\,(\ve{\sigma}-\ve{k}),
\qquad
\overline{D}^i=\left(\Lambda^i_a-\sigma^i\,\Lambda^0_a\right)\,D^a.
\end{eqnarray}

\noindent
Note that for the transformation of $\ve{R}=\ve{X}_p(T)-\ve{X}_A$ the
Lorentz transformation should be applied for the following two pair of
events: (1) $(t,x^i_p(t))$ and $(T,X^a_p(T))$, and (2) $(t,x^i_A(t))$
and $(T^\prime,X^a_A)$. Since the position $\ve{X}_A$ of the body in
the reference system $(T,X^a)$ is time-independent it is allowed to use
any moment $T^\prime$ to represent its position $(T^\prime,X^a_A)$
relative to that reference system. The moment $T^\prime$ is chosen in
such a way that the event $(t,x^i_A(t))$ corresponding to
$(T^\prime,X^a_A)$ has in the reference system $(t,x^i)$ the same time
coordinate as the event $(t,x^i_p(t))$. Using this scheme one easily
derives that $R^a=\overline{\Lambda}^a_i\,r^i$.

\section{Expressions through the position of the body
in retarded moment of time}
\label{appendix-retarded}

The retarded moment of time $t_*$ is a function of $t$ and the
positions of the photon $\ve{x}_p$ and the body $\ve{x}_A$ and is
related to these quantities through the null cone equation

\begin{equation}\label{t-*-t}
t_*+{1\over c}\,|\ve{r}_*|=t,
\end{equation}

\noindent
with

\begin{equation}\label{r_*}
\ve{r}_*=\ve{x}_{p0}+c\,\ve{\sigma}\,(t-t_0)-\ve{x}_A(t_*).
\end{equation}

\noindent
The unperturbed coordinates of the photon are sufficient in (\ref{r_*})
since we work in the first post-Newtonian approximation. For the case
of a body moving with a constant velocity an exact relation between
$\ve{r}$ and $\ve{r}_*$ can be derived. Using (\ref{r}), (\ref{r_*}),
(\ref{t-*-t}), (\ref{x_A(t)}) and
$\ve{x}_A(t)=\ve{x}_A(t_*)+\ve{v}_A\,(t-t_*)$ one gets (see Section
14.1 of \cite{Jackson:1974} for a geometric interpretation)

\begin{equation}\label{r-r_*}
\ve{r}=\ve{r}_*-|\ve{r}_*|\,\ve{k}.
\end{equation}

\noindent
Using $\ve{r_*}$ instead of $\ve{r}$ in the equations given in Appendix
\ref{Appendix-Lorentz-transformed} one gets
\begin{eqnarray}
\label{tr-abs-R-*}
|\ve{R}|&=&\gamma\,\left(|\ve{r}_*|-\ve{k}\cdot\ve{r}_*\right),
\\
\label{tr-R-SigmaR-*}
|\ve{R}|-\Sigma\cdot\ve{R}&=&{1\over \gamma\,\theta}\,
\left(|\ve{r}_*|-\ve{\sigma}\cdot\ve{r}_*\right),
\\
\label{tr-P-D_*}
\overline{\ve{D}}&=&
\ve{d}_*
-\ve{\sigma}\times(\ve{k}\times\ve{\sigma})\,\theta^{-1}\,
\left(|\ve{r}_*|-\ve{\sigma}\cdot\ve{r}_*\right),
\end{eqnarray}

\noindent
where $\theta$ is defined by (\ref{theta}) and
\begin{eqnarray}\label{d_*}
\ve{d}_*&=&\ve{\sigma}\times(\ve{r}_*\times\ve{\sigma}).
\end{eqnarray}

\section{The Kopeikin-Sch\"afer solution}
\label{appendix:Kopeikin:Schaefer}

An exact post-Minkowskian solution for the light propagation in the
gravitational field of moving mass monopoles has been first derived by
\citet{Kopeikin:Schaefer:1999}. The authors do not write, however,
their solution in explicit form needed for the purposes of this paper.
Taking Eqs. (10), (26), (30), (31), (34) and (35) from
\citet{Kopeikin:Schaefer:1999} one can write the Kopeikin-Sch\"afer
solution in the following explicit from:
\begin{eqnarray}\label{Delta-x-pM}
\Delta\ve{x}_p(t)&=&
-{2GM\over c^2}\,\biggl(\ve{f}^{KS}(t)-\ve{f}^{KS}(t_0)+\ve{g}(t_0,t)\biggr)+{\cal O}(G^2),
\\ \label{F-pM}
\ve{f}^{KS}(t)&=&\gamma_*\left(\theta_*\,{\ve{d}_*\over p_*}
-(\ve{\sigma}-\ve{k}_*)\,\log p_*\right),
\\ \label{cal-R-pM}
\ve{g}(t_0,t)&=&\int_{t_0}^t {\gamma_*^3\over c\,(|\ve{r}_*|-\ve{k}_*\cdot\ve{r}_*)}\,
\biggl[\,
\left(
\gamma^{-2}_*\,(\ve{\sigma}\cdot\ve{a}_*)
-\theta_*\,(\ve{k}_*\cdot\ve{a}_*)
\right)\,
\ve{d}_*
\nonumber\\
&&\phantom{\int_{t_0}^t \biggl[\,}
+p_*\,
\log p_*\,
\left(
(\ve{k}_*\cdot\ve{a}_*)
\,(\ve{\sigma}-\ve{k}_*)-\gamma^{-2}_*\,\ve{a}_*\right)
\biggr]\,dt,
\nonumber
\\
\\
\label{Delta-v-pM}
{1\over c}\,\Delta\ve{v}_p(t)&=&
-{2GM\over c^2}\,{\gamma_*\,\theta_*\over |\ve{r}_*|-\ve{k}_*\cdot\ve{r}_*}\,
\nonumber
\\
&&\times\,
\biggl(\theta_*\,{\ve{d}_*\over p_*}
+(2-\theta_*)\,\ve{\sigma}-2\,\ve{k}_*\biggr)
+{\cal O}(G^2),
\\
\label{a_*}
\ve{a}_*&=&\ddot{\ve{x}}_A(t_*),
\\
\label{gamma-*}
\gamma_*&=&\left(1-|\ve{k}_*|^2\right)^{-1/2},
\\
\label{theta-*}
\theta_*&=&1-\ve{\sigma}\,\cdot\,\ve{k}_*,
\\
\label{k_*}
\ve{k}_*&=&\ve{k}(t_*),
\end{eqnarray}

\noindent
where $p_*$ is defined by (\ref{p-*}), $\ve{d_*}$ by (\ref{d_*}) and
$\ve{r}_*$ by (\ref{r_*}). \citet{Kopeikin:Schaefer:1999} showed that
the integral $g(t,t_0)$ can be rewritten as a function of the retarded
times $t_*$ and $t_{0*}$ corresponding to the moments $t$ and $t_0$, so
that the null cone equation is not required while computing $g(t,t_0)$.
However, this is not important for the present paper and we prefer to
retain $g(t,t_0)$ in the form given by (\ref{cal-R-pM}).

Note that the Kopeikin-Sch\"afer solution is valid for arbitrarily
moving bodies and that the velocity $\ve{v}$ of the body and the vector
$\ve{k}=\ve{v}/c$ are not necessarily constants. The velocity $\ve{v}$
should be computed for the retarded moment $t_*$ in
(\ref{gamma-*})--(\ref{k_*}). The actual trajectory of the body should
be used in (\ref{t-*-t}) and (\ref{r_*}) to  compute $t_*$ and
$\ve{r}_*$.


\begin{thebibliography}{}

\bibitem[Bienayme \& Turon (2002)]{Bienayme:Turon:2002}
Bienaym\'e, O. \& Turon, C. eds. 2002,
GAIA: A European Space Project (Les Ulis: EDP Sciences)

\bibitem[ESA (2000)]{GAIA:2000}
ESA. 2000 GAIA: Composition, Formation and Evolution of the Galaxy,
Concept and Technology Study Report (ESA-SCI[2000]4) (Noordwijk: ESA)

\bibitem[Hellings (1986)]{Hellings:1986}
Hellings, R. W. 1986, \aj, 91, 650

\bibitem[Jackson (1974)]{Jackson:1974}
Jackson, J. D. 1974, Classical Electrodynamics (New York: Wiley)

\bibitem[Klioner (1989)]{Klioner:1989}
Klioner, S. A. 1989, Soobschch. Inst. Prik. Astron. No 6
(Communications of the Institute of Applied Astronomy, No 6, St.
Petersburg, in Russian)

\bibitem[Klioner (1991)]{Klioner:1991}
Klioner, S. A. 1991,
in: Geodetic VLBI: Monitoring Global Change, ed. W.E. Carter, NOAA
Technical Report NOS 137 NGS 49, 188 (Washington D.C.: American
Geophysical Union)

\bibitem[Klioner (2003)]{Klioner:2003}
Klioner, S. A. 2003, \aj, 125, 1580

\bibitem[Klioner \& Peip (2003)]{Klioner:Peip:2003}
Klioner, S. A., Peip, M. 2003,
Numerical Simulations of the Light Propagation in
Gravitational Field of Moving Bodies,
to be submitted to \aap

\bibitem[Klioner \& Kopeikin (1992)]{Klioner:Kopeikin:1992}
Klioner, S. A., Kopeikin, S. M. 1992, \aj, 104, 897

\bibitem[Kopeikin (1997)]{Kopeikin:1997}
Kopeikin, S. M. 1997, J. Math. Phys., 38, 2587

\bibitem[Kopeikin \& Sch\"afer (1999)]{Kopeikin:Schaefer:1999}
Kopeikin, S. M., Sch\"afer, G. 1999, \prd, 60, No. 124002

\bibitem[Kopeikin \& Mashhoon (2002)]{Kopeikin:Mashhoon:2002}
Kopeikin, S. M., Mashhoon, B. 2002, \prd, 65, No. 64025

\bibitem[Perryman et al. (2001)]{Perryman:et:al:2001}
Perryman, M.A.C., et al., 2001, \aap, 369, 339

\bibitem[Shao (1998)]{SIM:1998}
Shao, M., 1998, \procspie, 3350, 536

\bibitem[Will (1993)]{Will:1993}
Will, C. M. 1993,
Theory and experiment in gravitational physics,
(rev. ed.; Cambridge: Cambridge University Press)

\end{thebibliography}
\end{document}